\documentclass[a4paper,10pt]{article}
\usepackage[utf8x]{inputenc}

\usepackage{amssymb}
\usepackage{amssymb}
\usepackage{chemarr}
\usepackage{setspace}
\usepackage{lipsum,amsmath,multicol}
\usepackage{widetext}
\usepackage{pstricks}
\usepackage{pstricks-add}
\usepackage{xkeyval}
\usepackage{pst-plot}
\usepackage{pst-func}
\usepackage{subfigure}
\usepackage{tabularx}

\usepackage{graphicx}

\title{\huge \textbf{Survival of the Scarcer in Space}}
\author{\textbf{Renato Vieira dos Santos} \\ Departamento de F\'{i}sica, Instituto de Ci\^{e}ncias
Exatas, \\ Universidade Federal de Minas Gerais, CP 702, CEP 30161-970, \\ Belo Horizonte, Minas Gerais, Brasil.\\ e-mail:
\textit{econofisico@gmail.com} \\ \textbf{Ronald Dickman} \\ Departamento de F\'{i}sica, Instituto de Ci\^{e}ncias
Exatas, \\ and National Institute of Science and Technology for Complex Systems, \\
Universidade Federal de Minas Gerais, CP 702, CEP 30161-970, \\
Belo Horizonte, Minas Gerais, Brasil. }

\begin{document}

\maketitle

\begin{abstract}

The dynamics leading to extinction or coexistence of competing species
is of great interest in ecology and related fields.
Recently a model of intra- and interspecific competition between two species was proposed
by Gabel et al. [Phys. Rev. E 87 (2013) 010101], in which the
scarcer species (i.e., with smaller stationary population size) can be more resistant to
extinction when it holds a competitive advantage; the latter study considered
populations without spatial variation.
Here we verify this phenomenon in populations distributed in space.
We extend the model of Gabel et al. to a $d$-dimensional lattice, and study its population dynamics
both analytically and numerically.  Survival of the scarcer in space is verified
for situations in which the more competitive species is closer to the threshold for extinction
than is the less competitive species, when considered in isolation.
The conditions for survival of the scarcer species, as obtained applying renormalization
group analysis and Monte Carlo simulation,
differ in detail from those found in the spatially homogeneous case.
Simulations highlight the speed of invasion waves in determining the survival times
of the competing species.
\\


\end{abstract}


\section{Introduction}

Coexistence of species and maintenance of species diversity are key issues in ecology as well as in conservation and restoration.  A key idea in this context is
the competitive exclusion principle, which asserts that similar species competing for a limited resource cannot coexist \cite{competitive_exclusion}. 
Individual-based stochastic models, both with and without spatial structure, are useful for analyzing these questions \cite{RenshawBook1993}. There is some
evidence \cite{murrell2001uniting} to suggest that spatial structure should facilitate coexistence of similar competitors. This is because when dispersal
and interactions are localized, individuals tend to interact with conspecific neighbors more frequently than would be suggested by the overall densities at the
landscape level; this is a common pattern in natural communities \cite{condit2000spatial}. Computer simulations suggest that such spatial structures could lead
to long-term coexistence of similar competitors and hence to the maintenance of high levels of biodiversity \cite{silander1985neighborhood}. This mechanism has
been encapsulated in the segregation hypothesis \cite{pacala1995details}, which states that intraspecific spatial aggregation promotes stable coexistence by
reducing interspecific competition.

In the context of competition and extinction, the question of population size plays a fundamental role. In the simplest birth-and-death models, for example, the
extinction probability of a viable, but initially small population is high, but decreases rapidly with population size.\footnote{See Ch. 2 of
\cite{RenshawBook1993}.} Recently, Gabel, Meerson, and Redner \cite{survival_weaker} proposed a stochastic model of two-species competition in which the species
with smaller population size is, under certain conditions, less susceptible to extinction than the more populous species. This phenomenon, dubbed {\it survival
of the scarcer} (SS) in \cite{survival_weaker}, is somewhat surprising since conventional wisdom on population dynamics suggests that a smaller population is
more susceptible to extinction due to demographic fluctuations. The authors of \cite{survival_weaker} use a variant of the Wentzel-Kramers-Brillouin-Jeffreys
(WKBJ) approximation to obtain the tails of the quasi-stationary (QS) probability distribution \cite{assaf2006spectral,assaf2010extinction}. 
(Here and below, the quasi-stationary regime characterizes the behavior at long times, conditioned on survival.)
They show that asymmetry in interspecific competition can induce survival of the scarcer species.

While the analysis of \cite{survival_weaker} holds for well mixed (spatially uniform) populations, in view of the above mentioned segregation hypothesis
it is natural to inquire whether the same conclusions apply for populations distributed in space. In this paper we study a stochastic model similar to that of
\cite{survival_weaker}, but with organisms located on a $d-$dimensional lattice, and investigate the conditions required for survival of the scarcer in space
(SSS).  On reproduction, a daughter organism appears at the same site as the mother; competition only occurs between organisms at the same site. In addition to
these reactions, organisms also diffuse on the lattice.  We study the model using two complementary approaches: field-theoretic analysis and direct numerical
(Monte Carlo) simulation.

Since the field-theoretic study requires specialized techniques, we summarize the method for readers less familiar with this approach. Starting from the master
equation, we construct a representation involving creation and annihilation operators. This allows us to obtain a functional integral formulation, that is, a
mapping of the stochastic process to a field theory. This now standard procedure is known as the Doi-Peliti mapping \cite{Doi76a,peliti2,Tauber}. We consider
the population dynamics in a critical situation, i.e., competing populations close to extinction.  The dynamic renormalization group (DRG) is used to study the
nonequilibrium critical dynamics of the populations, specifically, to determine how the model parameters transform under rescaling of length and time.
The analysis furnishes a set of reduced stochastic evolution equations that describe the population dynamics for parameters close to those of a fixed point of
the DRG. Analysis of these equations lets us map out regions of coexistence and of single-species survival in the space of reproduction and competition rates. 
In particular, we find regions in which the species predicted by mean-field theory to have the larger population in fact goes extinct.

We use Monte Carlo simulations to probe the QS population densities and lifetimes of the two species.  We find SSS in a small but significant region of
parameter space, in which the less populous species is more competitive.  The simulations yield insights into the spatiotemporal pattern of population growth
and decay, the speed of species invasion, and the QS probability distribution.

The remainder of this paper is organized as follows. In section \ref{modelo} we describe the model and write the effective action associated with the stochastic
process. A brief analysis of mean-field theory, valid in dimensions greater than the critical dimension is performed. In section \ref{DRG} we use some known
results on multispecies directed percolation to obtain the renormalization group flow in parameter space. This analysis, combined with numerical simulations of
the associated stochastic partial differential equations, allows us to determine the regions in parameter space in which SSS occurs. Results of Monte Carlo
simulations are presented in Section \ref{simulation}. Section \ref{conclusion} closes the paper with our conclusions.

\section{Model}
\label{modelo}

The model proposed in \cite{survival_weaker} may be described, with some minor modifications in the rates as specified below, by the following set of reactions:

\begin{equation}
 \left\{\begin{matrix}
A\hspace{0.2cm}\overset{\alpha}{\rightharpoonup}\hspace{0.2cm}A+A \hspace{1cm}
B\hspace{0.2cm}\overset{\beta}{\rightharpoonup}\hspace{0.2cm}B+B\\
A+A\hspace{0.2cm}\overset{\alpha'}{\rightharpoonup}\hspace{0.2cm}\emptyset \ \ (A) \hspace{1cm}
B+B\hspace{0.2cm}\overset{\beta'}{\rightharpoonup}\hspace{0.2cm}\emptyset \ \ (B) \\
A+B\hspace{0.2cm}\overset{\zeta}{\rightharpoonup}\hspace{0.2cm}B \hspace{1cm}
A+B\hspace{0.2cm}\overset{\xi}{\rightharpoonup}\hspace{0.2cm}A
\end{matrix}\right.
\label{mod1}
\end{equation}

The reactions in the first line correspond to reproduction of species $A$ and $B$. Those in the second line describe intraspecific competition resulting in the
death of one individual ($A$/$B$) or of both ($\emptyset$), while the third line represents interspecific competition. $\alpha$ ($\beta$), $\alpha'$ ($\beta'$)
and $\zeta$ ($\xi$) are the rates of the reactions associated with species $A$ ($B$).

The relation between the rates in equation (\ref{mod1}) and those of the original model \cite{survival_weaker} is (in the case of mutual annihilation):
$\alpha\leftrightarrow1,$ $\beta\leftrightarrow g,$ $\alpha'\leftrightarrow 1/K,$ $\beta'\leftrightarrow 1/K,$ $\zeta\leftrightarrow\epsilon/K,$ and
$\xi\leftrightarrow\delta\epsilon/K.$ $ \delta<1$ implies an asymmetry in competition between species that favors $ B $ at the expense of $A.$ In this case
species $B$ is \textit{more competitive} than $A$; the opposite occurs if  $ \delta > 1.$

\begin{figure}
\centering
\begin{pspicture}(5,4.5)
\centering
\psaxes[linewidth=1.2pt,labels=none,ticks=none,axesstyle=frame]{->}(0,0)(0,0)(5,5)[\textbf{$\delta$},0][\textbf{$g$},90]

\psline[linewidth=1.5pt,linecolor=blue,linestyle=dashed](0,1.5)(5,3.5)
\psline[linewidth=1.5pt,linecolor=red](0,0.2)(5,4.8)

\rput(0.4,1.1){$I$}
\rput(3.5,0.8){$A$ dominant}
\rput(4.5,3.8){$II$}
\rput(1.4,4){$B$ dominant}

\qdisk(2.5,2.5){3pt}\uput[-90](2.5,2.5){}

\end{pspicture}\\[0.25cm]
\caption{Phase diagram of well mixed system in the $g$-$\delta$ plane, as furnished by the analysis
of Ref. \cite{survival_weaker}. Dashed blue curve:
$g_{mf}=(1+\delta\epsilon)/(1+\epsilon).$ Full red curve: $g_{ss}=(1+m\delta\epsilon)/(1+m\epsilon).$
}
\label{redner_fig}
\end{figure}
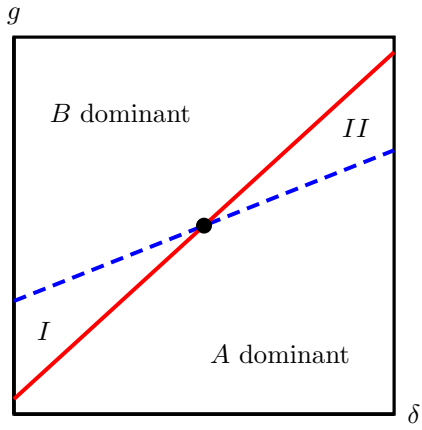

The authors of Ref. \cite{survival_weaker} constructed a phase diagram in the $ g $-$ \delta$ plane (shown schematically in Figure \ref{redner_fig}).
The phase diagram includes results of both mean-field theory (MFT) and analysis of the master equation for a stochastic population model in
a well mixed system, i.e., without spatial structure. In MFT, the populations of the two species are equal when
$g_{mf}(\delta)=(1+\delta\epsilon)/(1+\epsilon)$. Analysis of the master equation in the QS regime yields equal extinction probabilities when
$g_{ss}(\delta)=(1+m\delta\epsilon)/(1+m\epsilon)$ with $m\equiv\left[(\ln{2})^{-1}-1\right]^{-1}.$ These conditions are plotted in figure \ref{redner_fig};
they intersect at the point $(g,\delta)=(1,1).$ The phase diagram includes two ``normal" regions, in which the more populous species is less likely to go
extinct, and two ``anomalous" regions, in which the scarcer species is more likely to survive. Thus in region $I$ (where $\delta<1$ and $B$ is more competitive
in the sense defined above), species $A$ is more numerous in the quasi-stationary state and simultaneously more susceptible to extinction, while in region $II$
(where $\delta>1$ and $A$ is more competitive), species $B$ is more numerous, yet more likely to go extinct first. Regions $I$ and $II$ are anomalous since they
correspond to an inversion of the usual dictum that susceptibility to extinction grows with diminishing population size. Our goal in this work is to determine
whether a stochastic model of two-species competition with spatial structure, in which organisms diffuse in $d-$dimensional space, exhibits a similar
phenomenon. In the following subsection we develop a continuum description for this spatial stochastic process.

\subsection{Effective action}

We consider a stochastic implementation of the reactions of equation (\ref{mod1}) on a $d$-dimensional lattice, adding diffusion (nearest-neighbor hopping) of
individuals of species $A$ and $B$ at rates $D_A$ and $D_B$ respectively. Following the standard Doi-Peliti procedure mentioned in the Introduction, ignoring
irrelevant terms (in the renormalization group sense) and with the so-called Doi shift \cite{Tauber} already performed, we obtain the following effective
action:

\begin{widetext}
\begin{eqnarray}
 S(\bar{\phi},\phi,\bar{\psi},\psi) &=& \int d^dx\int
dt\left\{\bar{\phi}[\partial_t+D_A(\sigma_A-\nabla^2)]\phi-\alpha\bar{\phi}^2\phi+j\alpha'\bar{\phi}\phi^2+\alpha'\bar{\phi}
^2\phi^2+\zeta\psi\phi\bar{\phi}\right\} \nonumber \\
  & + & \int d^dx\int
dt\left\{\bar{\psi}[\partial_t+D_B(\sigma_B-\nabla^2)]\psi-\beta\bar{\psi}^2\psi+j\beta'\bar{\psi}\psi^2+\beta'\bar{\psi}
^2\psi^2+\xi\phi\psi\bar{\psi}\right\},
\label{mod4}
\end{eqnarray}
\end{widetext}
where $\sigma_A\equiv-\alpha/D_A,$ $\sigma_B\equiv-\beta/D_B$ and $j\equiv1$ ($2$) for individual death (mutual annihilation). Action $S$ has four fields. The
expected values of the fields $\phi$ and $\psi $ (evaluated by performing functional integrals over the four fields, using the weight $\exp[-S]$), represent the
mean population densities of species $A$ and $B$, respectively. The fields with overbars have no immediate physical interpretation, but are related to intrinsic
fluctuations of the stochastic dynamics. In particular, terms quadratic in $\overline{\phi}$ and $\overline{\psi}$ correspond to {\it noise}
in the associated stochastic evolution equations \cite{cardy}. The terms $\propto \phi \overline{\phi}^2$ and $\psi \overline{\psi}^2$
correspond, respectively, to intrinsic fluctuations of species A and B.  In the expansion of the action in powers of the fields, the lowest order term involving
the product $\overline{\phi}\overline{\psi}$ would be $\phi \psi \overline{\phi} \overline{\psi}$.  Since it is of fourth order in the fields, this term is {\it
irrelevant}, that is, the coefficient of this term flows to zero under repeated renormalization group transformations; it has therefore been excluded from the
action.  Fluctuations of one species nevertheless affect the other species via the coupling terms $\propto \zeta$ and $\xi$ in Eq. (\ref{mod4}).

Using the definitions of $\sigma_A$ and $\sigma_B$ and the change of variable $\bar{\phi}\to\theta_A\bar{\phi},$ $\phi\to\theta_A^{-1}\phi,$
$\bar{\psi}\to\theta_B\bar{\psi},$ and $\psi\to\theta_B^{-1}\psi,$ where $\theta_A\equiv\sqrt{\frac{j\alpha'}{\alpha}},$
$\theta_B\equiv\sqrt{\frac{j\beta'}{\beta}},$ we can write the effective action in the form:

\begin{eqnarray}
 S(\bar{\phi},\phi,\bar{\psi},\psi) &=& \int d^dx\int
dt\Big\{\bar{\phi}[\partial_t+D_A(\sigma_A-\nabla^2)]\phi  \nonumber \\
  & + & \bar{\psi}[\partial_t+D_B(\sigma_B-\nabla^2)]\psi  \nonumber \\
  & + & g_A\phi\bar{\phi}(\phi-\bar{\phi})+g_B\psi\bar{\psi}(\psi-\bar{\psi}) \nonumber \\
  & + & h_B\phi\psi\bar{\psi}+h_A\phi\psi\bar{\phi}  \Big\},
\label{mod7}
\end{eqnarray}
with
\begin{equation}
\left\{\begin{array}{rcl}
g_A&\equiv&\sqrt{j\alpha\alpha'} \\
h_A&\equiv&\theta_A^{-1}\zeta \\
g_B&\equiv&\sqrt{j\beta\beta'}\\
h_B&\equiv&\theta_B^{-1}\xi.
\end{array}\right.
\label{mod8}
\end{equation}
Dimensional analysis yields,

$$
[h_A]=[h_B]=p^{2-\frac{d}{2}}=[g_A]=[g_B],
$$
where $[X]$ denotes the dimensions of $X$, and $ p $ has dimensions of momentum \cite{Tauber}. The upper critical dimension is therefore $d_c=4$, as for
single-species processes such as the contact process and directed percolation.

\subsection{Mean-field approximation}

For $d \geq d_c, $ mean-field analysis, which ignores fluctuations in $\phi$ and $\psi,$ yields the correct critical behavior. Imposing the conditions
$\frac{\delta S}{\delta\phi}=0,$ $\frac{\delta S}{\delta\bar{\phi}}=0,$ $\frac{\delta S}{\delta\psi}=0$ and $\frac{\delta S}{\delta\bar{\psi}}=0,$ we obtain the
mean-field equations

\begin{equation}
\left\{\begin{array}{rcl}
\frac{\partial\phi}{\partial t}&=&D_A\nabla^2\phi+\alpha\phi-g_A\phi^2-h_B\phi\psi \\
\frac{\partial\psi}{\partial t}&=&D_B\nabla^2\psi+\beta\psi-g_B\psi^2-h_A\phi\psi.
\end{array}\right.
\label{MF}
\end{equation}
Let $L$ be a typical length scale and define $X=g_A\phi/\alpha,$ $Y=g_B\psi/\beta,$ $s=D_At/L^2$, and define a rescaled coordinate $x'=x/L$.  Then equation
(\ref{MF}) can be written in dimensionless form as

\begin{equation}
\left\{\begin{array}{rcl}
\frac{\partial X}{\partial s}&=&\nabla^2X+\gamma\left(X-X^2-aXY\right)\equiv\nabla^2X+f(X,Y) \\
\frac{\partial Y}{\partial s}&=&D\nabla^2Y+\gamma\left(cY-cY^2-bXY\right)\equiv D\nabla^2Y+g(X,Y)
\end{array}\right.
\label{MF2}
\end{equation}
with $\gamma\equiv\alpha L^2/D_A,$ $a\equiv \beta h_B/(\alpha g_B),$ $b\equiv\beta h_A/(\alpha g_B),$ $c\equiv g_A\beta^2/(g_B\alpha^2),$
$D\equiv D_B\beta g_A/(D_A\alpha g_B),$ $f(X,Y)\equiv\gamma\left(X-X^2-aXY\right)$ and $g(X,Y)\equiv\gamma\left(cY-cY^2-bXY\right).$
Coexistence is possible in the stationary state if the conditions $c>b$ and $a<1$ hold, implying $h_A<\beta g_A/\alpha$ and $h_B<\alpha
g_B/\beta.$ These conditions depend monotonically on parameters $\alpha,$ $\beta,$ $g_A$ and $g_B,$ analogous to the coexistence conditions found in
\cite{survival_weaker}. The same is not true in spatial stochastic theory, as we shall see in the next section.

\section{Renormalization group (RG) flow}
\label{DRG}

In this section we apply a renormalization group analysis to determine regions of coexistence and of single-species survival in the space of
reproduction and competition rates. Some years ago, Janssen analyzed a class of reactions of the kind defined in equation (\ref{mod1})
\cite{janssen2001directed}. This work considered multi-species reactions of the form

\begin{equation}
 X_{i}\leftrightarrow 2X_{i} \hspace{1cm} X_{i}\rightarrow\emptyset \hspace{1cm} X_i+X_j\rightarrow k X_i+l X_j,
\label{MPD}
\end{equation}
where $i$ and $j$ are species indices and $k,$ $l$ are either zero or unity; this process is called \textit{multicolored directed percolation} (MDP), with
different colors referring to different species. The RG analysis of \cite{janssen2001directed} shows that knowledge of the directed percolation fixed points is
sufficient to determine the fixed points of the full MDP. As shown in that work, the parameter combinations $g_A/D_B\equiv u_A$ and $g_B/D_A\equiv u_B,$ related
to intraspecific competition, flow under renormalization group transformations to the stable DP fixed point  $u_A^*=u_B^*=u^*=2\epsilon/3$ with
$\epsilon=4-d>0.$\footnote{Not to be confused with parameter $\epsilon$ used in equation \ref{mod1}.} Therefore, regarding the renormalization of intraspecific
competition parameters, inclusion of other species behaving as in DP does not alter the fixed point.

Janssen's analysis shows that there are four fixed points for interspecific competition parameters $h_A/D_A\equiv v_A$ and $h_B/D_B\equiv v_B,$ which from
reactions (\ref{mod1}) are related to $\xi$ and $\zeta$. The first two are $(v_A^*,v_B^*)=(0,0),$ which is unstable, and

\begin{equation}
 (v_A^*,v_B^*)=\left(\frac{D_A+D_B}{D_A}\frac{2}{3}\epsilon,\frac{D_A+D_B}{D_B}\frac{2}{3}\epsilon\right)
\label{MPD2}
\end{equation}
which is a hyperbolic fixed point if $D_A=D_B.$\footnote{In the nomenclature of \cite{janssen2001directed}, if species are of the same flavor.}
For $D_A\neq D_B,$ it was conjectured \cite{janssen2001directed} that the stability of the fixed points remains the same despite small changes in the
renormalization group flow diagram topology. The other two fixed points for $D_A=D_B$ are stable; they are given by $(v_A^*,v_B^*)=(0,2u^*)$ and
$(v_A^*,v_B^*)=(2u^*,0).$ To summarize, for $D_A=D_B\equiv D_0,$ the interspecies competition parameters flow to one of the following four $d-$dependent fixed
points (denoted as \textbf{O},\textbf{F},\textbf{G}, and \textbf{H} below),

\begin{eqnarray}
 (v_A,v_B)&=&\left(\frac{\theta_A^{-1}\zeta}{D_0},\frac{\theta_B^{-1}\xi}{D_0}\right) \nonumber \\
 &\to&\Big\{\textbf{O}:(0,0), \  \ \textbf{F}:(u^*,u^*), \nonumber \\
 & &  \  \ \textbf{G}:(2u^*,0), \  \ \textbf{H}:(0,2u^*)\Big\}
\label{FP}
\end{eqnarray}

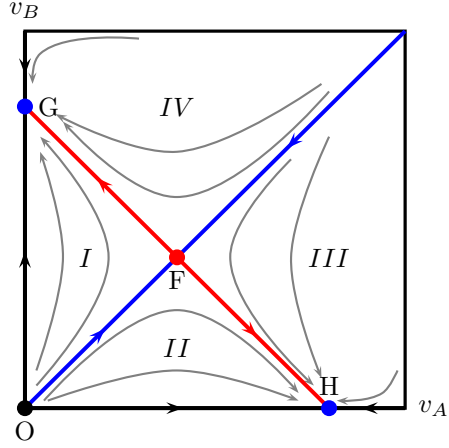
\begin{figure}
\centering
\begin{pspicture}(6,6)
\centering
\psaxes[linewidth=1.2pt,labels=none,ticks=none,axesstyle=frame]{->}(0,0)(0,0)(5,5)[\textbf{$v_A$},0][\textbf{$v_B$},90]

\pscurve[showpoints=false,linecolor=gray]{->}(0.15,0.3)(1.1,2)(0.2,3.6)
\pscurve[showpoints=false,linecolor=gray]{->}(0.15,0.5)(0.5,2)(0.2,3.4)
\pscurve[showpoints=false,linecolor=gray]{->}(0.25,0.1)(2,1.3)(3.6,0.2)
\pscurve[showpoints=false,linecolor=gray]{->}(0.3,0.1)(2,0.5)(3.6,0.1)
\pscurve[showpoints=false,linecolor=gray]{->}(3.5,3.3)(2.7,2)(3.8,0.3)
\pscurve[showpoints=false,linecolor=gray]{->}(4,3.6)(3.5,2)(3.9,0.4)
\pscurve[showpoints=false,linecolor=gray]{->}(4,4.2)(2,2.8)(0.5,3.8)
\pscurve[showpoints=false,linecolor=gray]{->}(3.9,4.3)(2,3.4)(0.4,3.9)
\pscurve[showpoints=false,linecolor=gray]{->}(1.5,4.9)(0.2,4.7)(0.1,4.3)
\pscurve[showpoints=false,linecolor=gray]{->}(4.9,0.5)(4.7,0.2)(4.1,0.1)
\psline[ArrowInside=->,linewidth=1.5pt,linecolor=blue](2,2)
\psline[ArrowInside=->,linewidth=1.5pt,linecolor=blue](5,5)(2,2)
\psline[ArrowInside=->,linewidth=1.5pt,linecolor=red](2,2)(0,4)
\psline[ArrowInside=->,linewidth=1.5pt,linecolor=red](2,2)(4,0)
\psline[ArrowInside=->,linewidth=1.5pt,linecolor=black](4,0)
\psline[ArrowInside=->,linewidth=1.5pt,linecolor=black](0,4)
\psline[ArrowInside=->,linewidth=1.5pt,linecolor=black](5,0)(4,0)
\psline[ArrowInside=->,linewidth=1.5pt,linecolor=black](0,5)(0,4)
\qdisk(0,0){3pt}\uput[-90](0,0){O}

\rput(0.8,2){$I$}
\rput(2,0.8){$II$}
\rput(4,2){$III$}
\rput(2,4){$IV$}
\psset{linecolor=red}
\qdisk(2,2){3pt}\uput[-90](2,2){F}
\psset{linecolor=blue}
\qdisk(0,4){3pt}\uput[-0](0,4){G}
\qdisk(4,0){3pt}\uput[90](4,0){H}

\end{pspicture}\\[0.25cm]
\caption{RG flow of the interspecific competition parameters $v_B$ and $v_A.$ Only the first quadrant is relevant due to the requirement of positive rates.
There is an unstable fixed point \textbf{O} at $(v_A,v_B)=(0,0)$ (black dot); a hyperbolic fixed point \textbf{F} at $(v_A,v_B)=(u^*,u^*)$ (red dot), and two
stable fixed points \textbf{G} and \textbf{H}, located at $(v_A,v_B)=(0,2u^*)$ and $(v_A,v_B)=(2u^*,0)$ respectively (blue dots). The separatrix $v_A=v_B$
is the boundary between the basins of attraction of \textbf{G} and \textbf{H}.}
\label{FLOW}
\end{figure}

Figure \ref{FLOW} shows the renormalization group flow diagram. The unstable fixed point \textbf{O} corresponds to complete decoupling between the two species,
and the hyperbolic point \textbf{F} to symmetric coupling. In the symmetric subspace, any nonzero initial value of $v \equiv v_A = v_B,$ flows to \textbf{F}.
For $v_B > v_A$ ($\delta>1$), $(v_A,v_B)$ flows to \textbf{G}, so that species $A$ is effectively unable to compete with $B$. Similarly, if $v_A > v_B$
($\delta<1$), the flow attains point \textbf{H}. In the following subsection we discuss the stationary states associated with these fixed points.

\subsection{Steady state close to critical points}

Population dynamics in the critical regime is governed by a pair of coupled stochastic partial differential equations (SPDE), which are readily deduced from the
action of equation (\ref{mod7})\cite{Tauber}:

\begin{equation}
 \frac{\partial\phi}{\partial t}=D_0\nabla^2\phi+\alpha\phi-g_A\phi^2-h_B\phi\psi+\eta_1
\label{cd1}
\end{equation}
and

\begin{equation}
 \frac{\partial\psi}{\partial t}=D_0\nabla^2\psi+\beta\psi-g_B\psi^2-h_A\psi\phi+\eta_2
\label{cd2}
\end{equation}
where the noise terms $\eta_1({\bf{x}},t)$ and $\eta_2({\bf{x}},t)$ satisfy
$\langle\eta_1({\bf{x}},t)\rangle =\langle\eta_2({\bf{x}},t)\rangle=0,$ and

\begin{equation}
 \langle\eta_1({\bf{x}},t)\eta_1({\bf{x}},t)\rangle=2g_A\phi({\bf{x}},t)\delta^d({\bf{x}}-{\bf{x}'})\delta(t-t'),
\label{cd3}
\end{equation}

\begin{equation}
 \langle\eta_2({\bf{x}},t)\eta_2({\bf{x}},t)\rangle=2g_B\psi({\bf{x}},t)\delta^d({\bf{x}}-{\bf{x}'})\delta(t-t').
\label{cd4}
\end{equation}
Note that these multiplicative noise terms depend on the square root of their respective fields, and were obtained directly from the action, without any
additional hypotheses.

\subsubsection{SPDE in the vicinity of point \textbf{F}}

In the symmetric subspace, $ \delta = 1,$ the RG flow is to the fixed point $ {\bf{F}},$ and parameters $g_A/D_0,$  $ g_B/D_0,$ $h_A/D_0,$ and $h_B/D_0$ take
the associated values. Therefore the SPDEs in (\ref{cd1}) and (\ref{cd2}) can be written as \cite{DRG}:

\begin{equation}
 \frac{\partial\phi}{\partial t}=D_0\nabla^2\phi+\phi-D_0u^*\phi^2-D_0u^*\phi\psi+\eta_1,
\label{cd5}
\end{equation}

\begin{equation}
 \frac{\partial\psi}{\partial t}=D_0\nabla^2\psi+g\psi-D_0u^*\psi^2-D_0u^*\psi\phi+\eta_2
\label{cd6}
\end{equation}
where we set $\alpha=1$ and $\beta=g.$ With the rescaling $\phi\to\phi/(D_0u^*),$ $\psi\to\psi/(D_0u^*)$ and $x\to(1/D_0)^{1/2}x,$ we can write equations
(\ref{cd5}) and (\ref{cd6}) in the form:

\begin{equation}
 \frac{\partial\phi}{\partial t}=\nabla^2\phi+\phi-\phi^2-\phi\psi+\eta_1,
\label{cd7}
\end{equation}
\begin{equation}
 \frac{\partial\psi}{\partial t}=\nabla^2\psi+g\psi-\psi^2-\psi\phi+\eta_2
\label{cd8}
\end{equation}
with rescaled noise
\begin{equation}
 \langle\eta_1({\bf{x}},t)\eta_1({\bf{x}},t)\rangle=2\left(D_0\right)^{2-\frac{d}{2}}(u^*)^2\phi({\bf{x}},t)\delta^d({\bf{x}}-{\bf{x}}
')\delta(t-t')
\label{cd9}
\end{equation}
\begin{equation}
\langle\eta_2({\bf{x}},t)\eta_2({\bf{x}},t)\rangle=2\left(D_0\right)^{2-\frac{d}{2}}(u^*)^2\psi({\bf{x}},t)\delta^d({\bf{x}}
-{\bf{x}}')\delta(t-t').
\label{cd10}
\end{equation}

\begin{figure*}
  \centering
  \mbox{
    \subfigure[Symmetric case with $\sigma=0.0$ and $g=1.$ Initial populations are $\phi (x, 0) = 0.4$ and $\psi (x, 0) =
0.6.$\label{sim2}]{\includegraphics[width=0.45\textwidth]{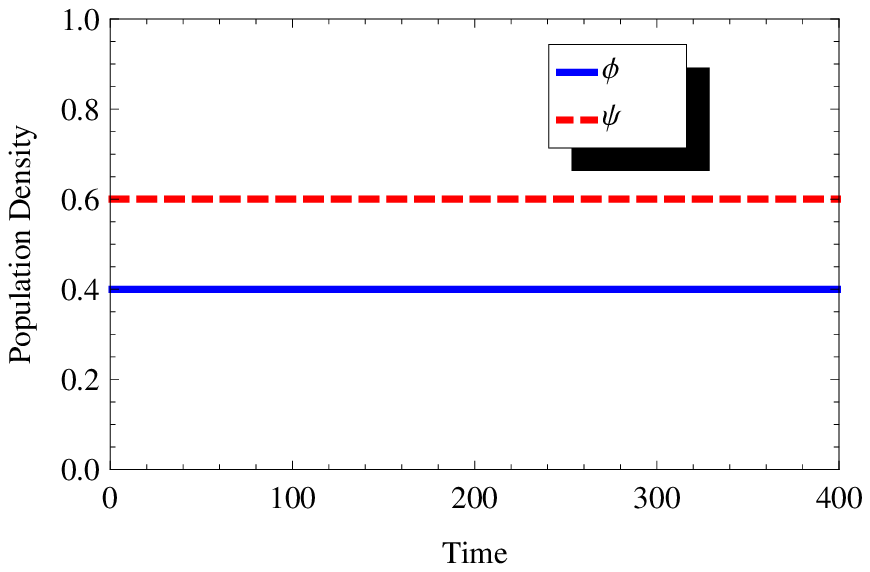}}\quad
    \subfigure[Symmetric case with $\sigma=0.03$ and $g=1.$ Initial populations are $\phi (x, 0) = 0.4$ and $\psi (x, 0) =
0.6.$\label{sim1}]{\includegraphics[width=0.45\textwidth]{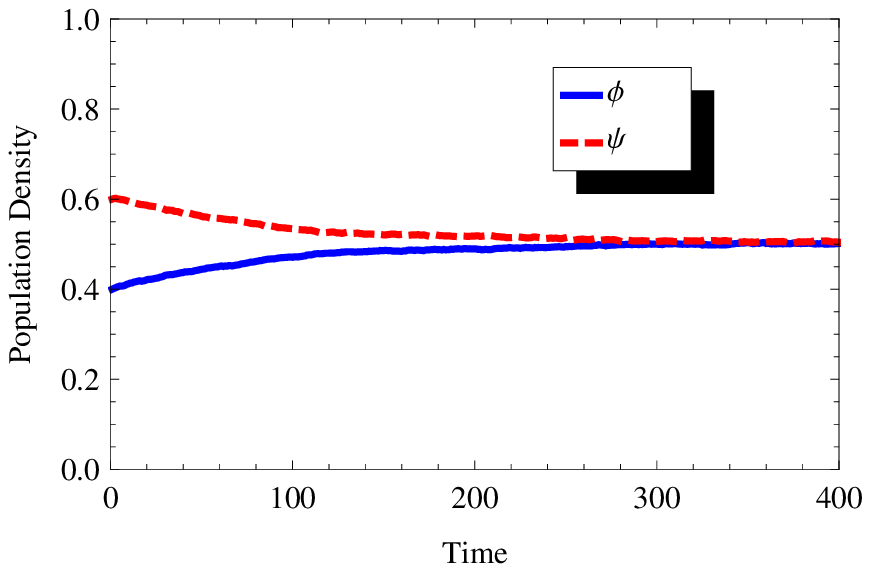}}\quad
   }
  \caption{Numerical simulations symmetric SPDE.}
  \label{two_figures}
\end{figure*}

For $d<4,$ the rescaled noise intensity $\sigma$ grows with diffusion rate $D_0;$ for $d=1$ we have $\sigma^2=2D_0^{\frac{3}{2}}u^{*2}$ with $u^{*2}\approx2.$
Figures \ref{sim2} and \ref{sim1} show results of numerical simulations of equations (\ref{cd7}) and (\ref{cd8}) in $d=1$ with $g = 1$. They show, for a
one-dimensional lattice of $L = 128$ sites in the interval $(-1,1),$ the population densities averaged over $1000$ Monte Carlo realizations for two different
values of the noise intensity. The time interval $ T $ is $ T= 400, $ partitioned into $N=40000$ steps so that $ \Delta t\equiv T/N=400/40000 = 0.01.$
Initial conditions are $\phi (x, 0)=0.4$ and $\psi (x, 0)=0.6.$ For $\sigma=0,$ population densities are almost time-independent as shown in Figure \ref{sim2}.
This is not the case for larger noise values. In this case, the two populations tend to the same value (see figure \ref{sim1}). Although Figs. \ref{sim2} and
\ref{sim1} represent averages over many realizations, we have verified coexistence in individual runs.

These simulations were performed using standard integration techniques for the Langevin equation with the XMDS2 software \cite{xmds2}. Since we are interested
in showing only the initial temporal trends of the two population densities in the vicinity of fixed points, the use of standard techniques of integrating the
Langevin equations is sufficient to reveal the qualitative nature of the solutions. Near a phase transition to an absorbing state, one of the population
densities tends to zero and the standard numerical integration scheme fails. This standard algorithm is not suitable for extracting the more accurate results
associated with critical exponents or asymptotic decays. In this case we would have to use more sophisticated algorithms such as those proposed in
\cite{dickman1994numerical,moro2004numerical,dornic2005integration}.

\subsubsection{SPDE in the vicinity of a stable fixed point}

Now consider the case $\delta\neq0.$ With $\delta>1,$ the parameters flow to point ${\bf{G}}$ and Eqs. (\ref{cd1}) and (\ref{cd2}) become

\begin{equation}
 \frac{\partial\phi}{\partial t}=D_0\nabla^2\phi+\phi-u^*\phi^2+\eta_1,
\label{cd11}
\end{equation}
\begin{equation}
 \frac{\partial\psi}{\partial t}=D_0\nabla^2\psi+g\psi-u^*\psi^2-2u^*\psi\phi+\eta_2,
\label{cd12}
\end{equation}
where we put $\alpha=1$ and $\beta=g.$ With a rescaling similar to that used above, we have

\begin{equation}
 \frac{\partial\phi}{\partial t}=\nabla^2\phi+\phi-\phi^2+\eta_1,
\label{cd13}
\end{equation}
\begin{equation}
 \frac{\partial\psi}{\partial t}=\nabla^2\psi+g\psi-\psi^2-2\phi\psi+\eta_2
\label{cd14}
\end{equation}
with $\eta_1$ and $\eta_2$ satisfying (\ref{cd9}) and (\ref{cd10}), respectively. For $\delta <1$ the situation is completely analogous to the case $\delta>1$,
and the resulting equations are as above with the exchange of $\phi$ and $\psi$ and changing the position of the factor $g$ accordingly.

\begin{figure}[htp]
\centering
\includegraphics[width=0.45\textwidth]{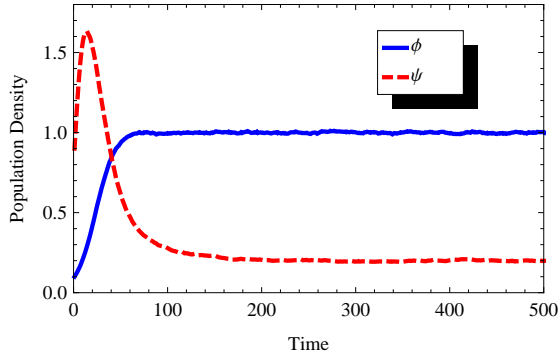}
\caption{Asymmetric case ($\delta=1.5$) with $\sigma=0.03$ and $g=2.2.$ Initial populations for $A$ and $B$ species are $\phi (x, 0) = 0.1$
and $\psi (x, 0) = 0.9,$ respectively.}\label{f2}
\end{figure}

If we neglect diffusion and noise terms, we have a system of  ordinary differential equations having a fixed point associated with the stable coexistence state
given by $(\phi^*,\psi^*)=(1,g-2).$\footnote{For the case $\delta <1,$ fixed points are $(\phi^*,\psi^*)=(1-2g,g),$ and coexistence condition is $0<g<1/2.$}
Therefore, the condition for coexistence is $ g > 2. $ If $ g<2,$ only species $A$ survives for $\delta>1.$ Figure (\ref{f2}) shows the result of a simulation
with $g=2.2$ in which the equations are integrated including both diffusion and noise. Numerical experiments indicate that the effect of these terms is to
increase the value of $ g\gtrapprox 2 $ for coexistence. Higher noise values imply higher thresholds $g$ for coexistence.

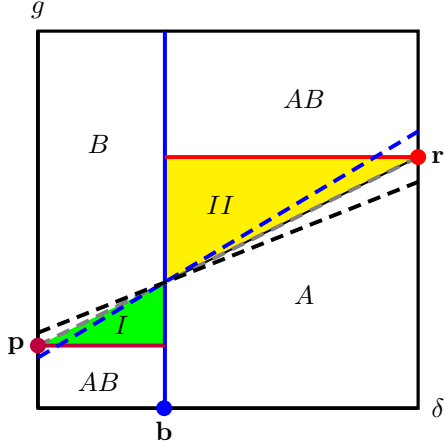
\begin{figure}
\centering
\begin{pspicture}(5,6)
\centering
\psaxes[linewidth=1.2pt,labels=none,ticks=none,axesstyle=frame]{->}(0,0)(0,0)(5,5)[\textbf{$\delta$},0][\textbf{$g$},90]

\pscustom[fillstyle=solid,fillcolor=green]{
\psline(0,0.83)(1.66,0.83)
\psline(1.67,0)(1.67,1.67)}

\pscustom[fillstyle=solid,fillcolor=yellow]{
\psline(1.67,3.33)(5,3.33)
\psline(5,3.33)(1.67,1.67)}

\psline[linewidth=1.5pt,linecolor=blue](1.67,0)(1.67,5)
\psline[linewidth=1.5pt,linecolor=purple](0,0.83)(1.67,0.83)
\psline[linewidth=1.5pt,linecolor=red](1.67,3.33)(5,3.33)
\psline[linewidth=1.5pt,linecolor=gray,linestyle=dashed](0,0.83)(5,3.33)
\psline[linewidth=1.5pt,linecolor=blue,linestyle=dashed](0,0.67)(5,3.67)
\psline[linewidth=1.5pt,linecolor=black,linestyle=dashed](0,1)(5,3)

\rput(0.8,3.5){$B$}
\rput(3.5,1.5){$A$}
\rput(3.5,4.1){$AB$}
\rput(0.8,0.35){$AB$}

\psset{linecolor=red}
\qdisk(5,3.33){3pt}\uput[0](5,3.33){\textbf{r}}
\psset{linecolor=blue}
\qdisk(1.66,0){3pt}\uput[-90](1.66,0){\textbf{b}}
\psset{linecolor=purple}
\qdisk(0,0.83){3pt}\uput[180](0,0.83){\textbf{p}}

\rput(1.1,1.1){$I$}
\rput(2.4,2.7){$II$}

\end{pspicture}\\[0.25cm]
\caption{Stationary phase diagram in the $ g $-$ \delta$ plane for stochastic model with spatial structure. The reference points have coordinates: 
${\bf{b}}=(1,0),$  ${\bf{p}}=(0,1/2),$ ${\bf{r}}=(3,2).$ For $ \delta> 1 $, when $ g < 2$ only species $ A $ is present, while for $g>2 $ there is coexistence.
Similarly, for $ \delta< 1 $, when $ g > 1/2$ only species $ B $ is present, and for $ g<1/2 $ there is coexistence. The diagonal line connecting points $
{\bf{p}} $ and $ {\bf{r}} $ is the equal-population criterion furnished by mean-field theory, $g_{mf}(\delta)=(1+\epsilon \delta)/(1+\epsilon)$, with $ \epsilon
= 1.$ In regions $I$ and $II,$ the more populous species ($A$ or $B$, respectively), as predicted by mean-field theory, in fact goes extinct. The thick, dashed,
blue and black curves represent $g_{mf}$ and $g_{sw}$ for the value of $ \epsilon $ that maximizes the area between them, i.e., $\epsilon=1/\sqrt{m}\approx
0.66.$}
\label{Phase_Diagram}
\end{figure}

For $\delta< 1$ similar reasoning applies. The results are summarized in the phase diagram of Fig. \ref{Phase_Diagram}. There are four stationary phases: a
pure-$A$ phase for $\delta > 1$ and $g<2,$ a symmetric pure-$B$ phase for $\delta<1$ and $g < 1/2,$ and two (disjoint) coexistence phases. The latter arise when
the less competitive species proliferates sufficiently faster than the more competitive one.

\subsubsection{Nature of phase transitions}

From the equations that omit the diffusion and noise terms, one may infer the nature of the phase transitions in the diagram of Fig. \ref{Phase_Diagram}. From
Table \ref{table}, which shows the fixed points of the equations for different values of $\delta,$ we see that in general the values in the final column do not
match for $\delta\neq1.$
\begin{table}[ht]
\caption{Simplified equations and their fixed points for different values of $\delta$}
\centering
\begin{center}
    \begin{tabular}{ | c | c | c |}
    \hline \hline
    $\delta$ & Equation & Fixed Point  \\ \hline \hline
    $=1$ & $\dot{\phi}=\phi-\phi^2-\phi\psi$ and $\dot{\psi}=g\psi-\psi^2-\phi\psi$ & $(\phi^*,\psi^*)=(0,g)$    \\ \hline
    $>1$ & $\dot{\phi}=\phi-\phi^2$ and $\dot{\psi}=g\psi-\psi^2-2\phi\psi$ & $(\phi^*,\psi^*)=(1,2-g)$  \\ \hline
    $<1$ & $\dot{\phi}=g\phi-\phi^2$ and $\dot{\psi}=\psi-\psi^2-2\phi\psi$ & $(\phi^*,\psi^*)=(g,1-2g)$  \\
    \hline \hline
    \end{tabular}
\label{table}
\end{center}
\end{table}
This fact indicates that the vertical blue line in the diagram (\ref{Phase_Diagram}), i.e., the line $\delta=1$, is a line of discontinuous phase transitions. 
Similarly, examining the expressions in the final column of Table \ref{table}, we infer the nature of phase transitions along the horizontal lines at $g = 1/2$
and $g = 2.$ In regions of coexistence, the densities of the minority species grow continuously from zero; thus these transitions are continuous. One should of
course recognize that the predictions for the phase diagram are essentially qualitative; quantitative predictions for nonuniversal properties such as phase
boundaries are not possible once irrelevant terms have been discarded.  Regarding the nature of the transitions, we expect the continuous ones, as furnished by
this mean-field-like analysis, to in fact belong to the DP universality class \cite{janssen2001directed}.  The line of discontinuous transitions contains a
portion (between levels p and r) that is between absorbing subspaces (only on species present in each phase), and other portions that separate (from the
viewpoint of the species undergoing extinction) an active and an absorbing phase.  Discontinuous transitions of this nature are not expected to occur in one
dimension \cite{hinrichsen1}.

According to mean-field theory, a point in region $ II $ corresponds to $ \rho_B> \rho_A $. If we increase the competitiveness of species $A$, increasing $
\delta $ while maintaining $ g $ fixed, there will be a point beyond which species $ B $ no longer has the greater population density. Improvement of the
competitiveness of species $A$ will make it the more populous species if organisms are well mixed.  In the case of a spatially structured population, by
contrast, {\it any} competitive advantage of species $A$, no matter how small, is sufficient to dominate the $B$ population (if the latter does not proliferate
very rapidly) when the population densities are very low, as is the case near criticality. Thus in a situation in which mean-field theory predicts species $B$
to be the majority, we can have species $A$ \textit{exclusively}. This can be seen as a strong version of the survival of the scarcer phenomenon.

To compare the predictions of the mean-field and spatially structured descriptions, a pair of dashed lines are plotted in Fig. \ref{Phase_Diagram}, representing
equations for $g_{mf}$ and $g_ {sw}$ as defined previously, for parameters such that the area between them is maximum, thereby maximizing the region in which SS
occurs. [The maximum area is obtained using $\epsilon = 1 / \sqrt {m}$ with $m=[(\ln{2})^{-1}-1]^{-1}$, in Eq. (\ref{mod1})]. Given the much larger area of the
region exhibiting SS in the spatial model, compared with that in the mean-field analysis (i.e., the area between the two dashed lines for $ 1<\delta<3,$) we see
that the SS phenomenon can be greatly intensified when spatial structure is included. Analogous reasoning applies for region $I$ (green). (For the parameter
values used, however, this reasoning does not apply for $\delta> 3$, since in this case the area between the dashed lines can be arbitrarily large. Also we no
longer have a region analogous to region $II$ with species $A$ only.) In this way we see that the phase diagram for the spatial stochastic model is rather
different from that found in \cite{survival_weaker}. Moreover, under certain conditions, the survival of the scarcer phenomenon is strengthened.


\section{Simulations}
\label{simulation}

Since the preceding analysis involves approximations whose reliability is difficult to assess, we perform simulations of a simple lattice model to verify SSS. 
This approach permits us to access the QS regime of the spatial model, in finite systems. We consider the following spatial stochastic process, defined on a
lattice of $L^d$ sites.  Each site $i$ is characterized by nonnegative occupation numbers $a_i$ and $b_i$ for the two species. The organisms (hereafter
``particles") of the two species evolve according to the following rates:

\begin{itemize}
 \item Particles of either species hop to a neighboring site at rate $D$.
 \item Particles of species A(B) reproduce at rate $\lambda_A$($\lambda_B$).
 \item At sites with two or more A particles, mutual annihilation occurs at rate $\alpha_A a_i (a_i-1)$,
 and similarly for B particles, at rate $\alpha_B b_i (b_i-1)$.
 \item At sites having both A and B particles, the competitive reaction B $\to$ 0 occurs at
 rate $\zeta_A a_i b_i$ and the reaction A $\to$ 0 occurs at rate $\zeta_B a_i b_i$.
\end{itemize}

The model is implemented in the following manner.  In each time step, of duration $\Delta t,$ each process is realized, at all sites, in the sequence: hopping,
reproduction, annihilation, competition.

In the hopping substep, each site is visited, and the probability of an A particle hopping to the right (under periodic boundaries) is set to $p_h =D a_i \Delta
t/(2d)$. A random number $z$, uniform on  $[0,1)$ is generated, and if $z < p_h$ the site is marked to transfer a particle to the right.  Once all sites have
been visited, the transfers are realized.  The same procedure is applied, in parallel, for B particles hopping to the right. Subsequently, hopping in the other
$2d-1$ directions is realized in the same manner.

In the reproduction substep, the A particle reproduction probability at each site $i$ is taken as $p_c = \lambda_A a_i \Delta t.$  A random number $z$ is
generated and a new A particle created at this site if $z < p_c.$  An analogous procedure is applied for reproduction of B particles.

In the annihilation substep, a probability $p_a = \alpha_A a_i (a_i-1) \Delta t$ is defined at each site, and mutual annihilation ($a_i \to a_i -2$) occurs if a
random number is $< p_a.$  Again, an analogous procedure is applied for annihilation of B particles.

Finally, at sites harboring both A and B particles, probabilities $p_A = \zeta_A a_i b_i \Delta t$ and $p_B = \zeta_B a_i b_i \Delta t$ are defined. The process
B $\to$ 0 occurs if a random number $z < p_A$, while the complementary process A $\to$ 0 occurs if $ p_A \leq z < p_A + p_B.$ Note that at most one of the
processes (A $\to$ 0 and B $\to$ 0) can occur at a given site in this substep.

The time step $\Delta t$ is chosen to render the reaction probabilities relatively small.  To do this, before each step we scan the lattice and determine the
maximum, over all sites, of $a_i,$ $b_i,$ and $a_i b_i.$ With this information we can determine the maximum reaction rate over all sites and all reactions.
Then $\Delta t $ is taken such that the maximum reaction probability (again, over sites and reactions) be 1/5. In this way, the probability of multiple
reactions (e.g., two A particles hopping from $i$ to $i+1,$ etc.) is at most $1/25,$ and can be neglected to good approximation, particularly in the regime in
which occupation numbers are typically small.

We simulate of the process on a ring ($d=1$) using {\it quasistationary} (QS) simulations, intended to sample the QS probability distribution
\cite{qssim,qssim2}.  In these studies the system is initialized with one A and one B particle at each site. When either species goes extinct (an absorbing
subspace for the process), the simulation is reinitialized with one of the active configurations (having nonzero populations for both species) saved during the
run. Following a brief transient, the particle densities fluctuate around steady values. We search for parameter values such that species A is more numerous,
despite being much less competitive than species B (i.e., $\zeta_B >> \zeta_A$).

Given the large parameter space, certain rates are kept fixed in the study: We set $D = \alpha_A = \alpha_B = 1/4.$ To understand the competitive
dynamics we first need to determine the conditions for {\it single-species} survival. We use QS simulations as well as spreading simulations \cite{torre} to
estimate the critical value of $\lambda$ for survival of single species, given $\alpha=1/4.$ In spreading simulations the initial configuration is that of
a single site with one particle, and all other sites empty.  We search for the value of $\lambda$ associated with a power-law behavior of the survival
probability $P(t)$ and the mean population size $n(t).$ These studies yield $\lambda_c = 1.267(1)$ as the critical point for survival of a single species,
i.e., without interspecies competition.  (Here and in the following, figures in parentheses denote statistical uncertainties in the final digit.)
The phase transition is clearly continuous; details on scaling behavior will be reported elsewhere.

\begin{figure}[!htb]
\centering
\includegraphics[trim=3cm 5cm 0.3cm 9cm, clip=true]{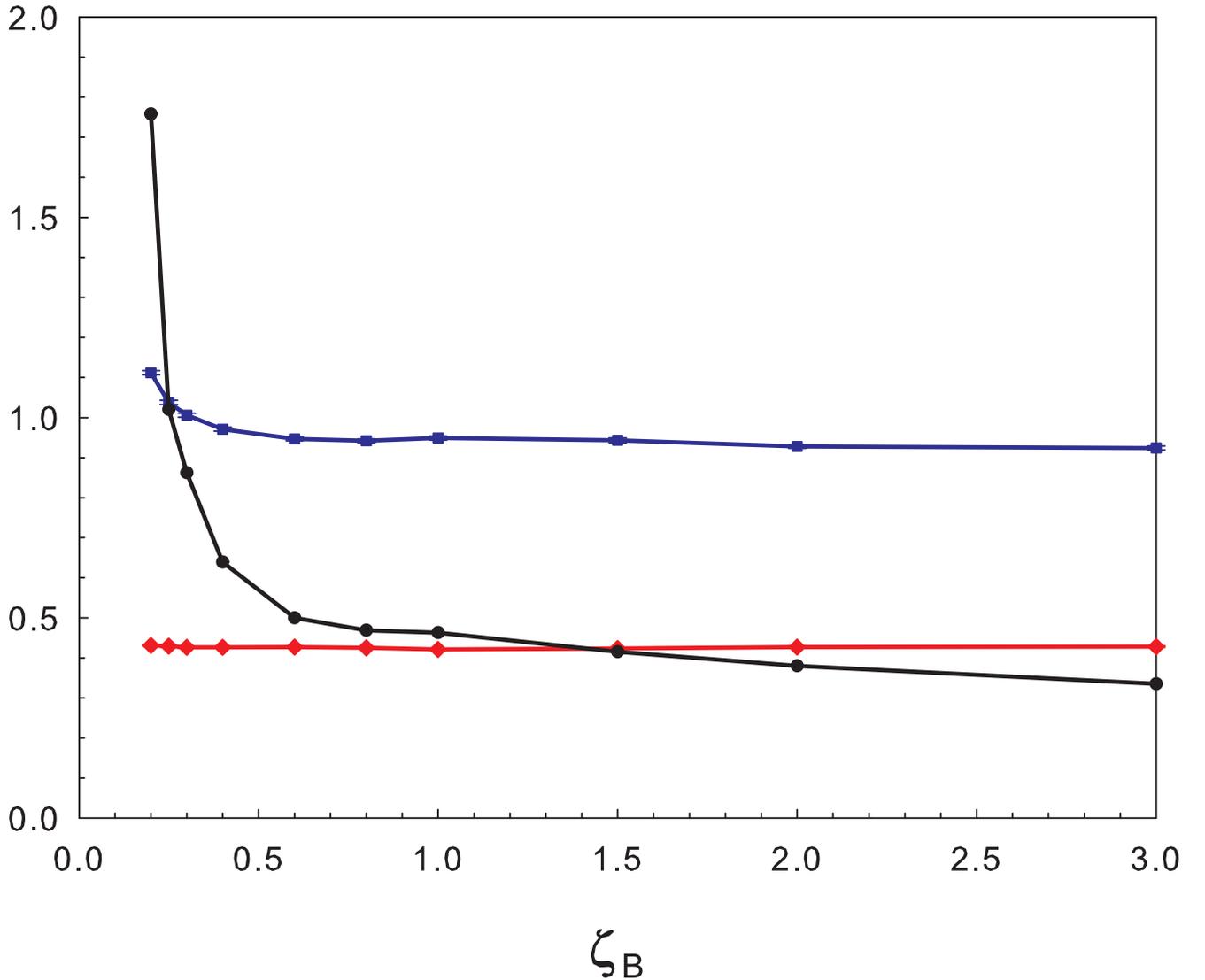}
\caption{Quasistationary population densities $\rho_A$ (blue) and $\rho_B$ (red), and lifetime ratio $\tau_A/\tau_B$ (black)
versus $\zeta_B$, for parameters as specified in text.}
\label{r125}
\end{figure}

\begin{figure}[!htb]
\centering
\includegraphics[trim=2cm 1cm 2cm 1cm, clip=true]{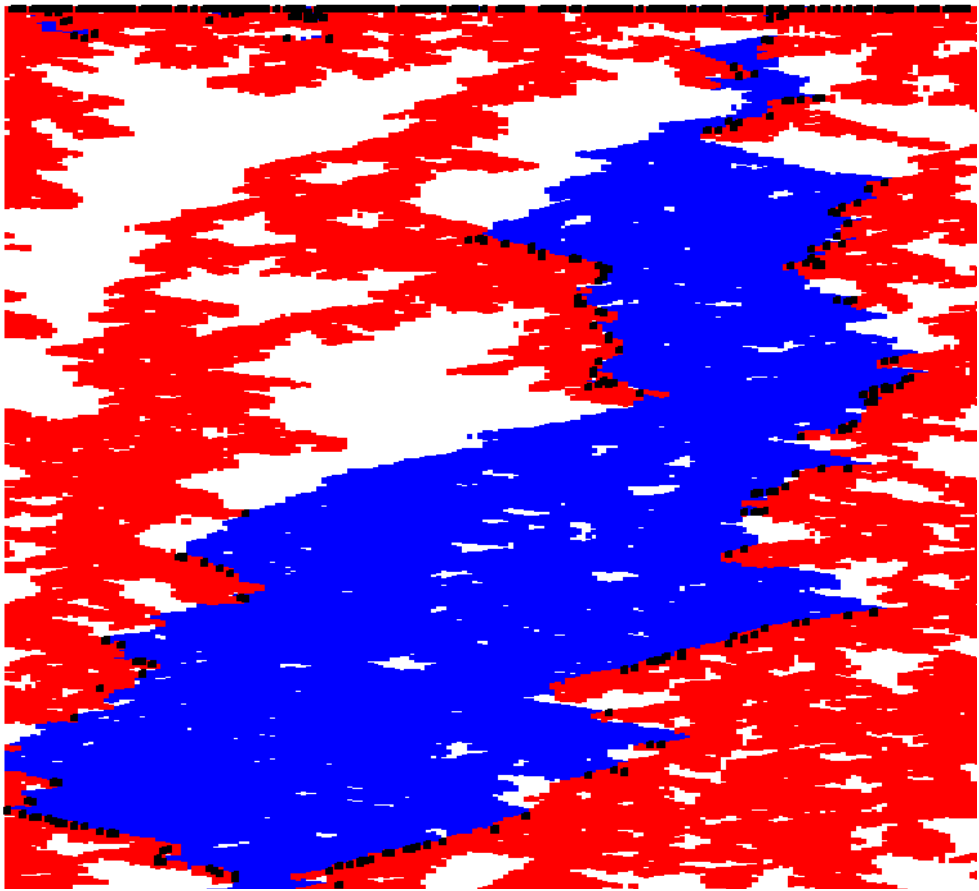}
\caption{A sample evolution on a ring of 200 sites, of duration 4000 time units (with time increasing downward). Blue: sites with $a_i > 0$
and $b_i=0$; red: sites with $a_i=0$ and $b_i>0$; black: sites with both species present.}
\label{ss5}
\end{figure}

In the two-species studies we set $\lambda_A = 1.6$ and $\zeta_A = 0.005,$ so that species A is well above criticality but weakly competitive.  We then vary
$\lambda_B$ and $\zeta_B$, monitoring the QS population densities $\rho_A$ and $\rho_B$ of the two species, as well as their QS lifetimes, $\tau_A$ and
$\tau_B.$ The latter are estimated by counting the number of times, in a long QS simulation, that one or the other species goes extinct. Survival of the scarcer
is then characterized by the conditions $\rho_A > \rho_B$ and $\tau_A < \tau_B.$ For the parameter values studied, we observe SSS for $\lambda_B \approx
\lambda_c$ and $\zeta_B \gg \zeta_A.$  Figure \ref{r125} illustrates the variation of the population densities, and of the ratio $\tau_A/\tau_B,$ as $\zeta_B$
is varied for fixed $\lambda_B = 1.25$, on a ring of 200 sites.  In this case, $\rho_A > \rho_B$ throughout the range of interest; the scarcer species, B,
survives longer than A for $\zeta_B \geq 0.25.$ Figure \ref{ss5} shows a typical evolution, for $\zeta_B=2$ and other parameters as in Fig. \ref{r125}.
(The spatiotemporal pattern observed for other parameter sets exhibiting SSS is qualitatively similar.) The system is divided into A-occupied regions,
B-occupied regions, and voids. The voids arising in B-occupied regions are larger, as this species is nearer criticality. Species A, which is well above
criticality, rapidly invades empty regions, but is in turn subject to invasion by the more competitive species B. Sites bearing both species are quite rare,
occurring only at the frontiers between regions occupied by a single species.

\begin{figure}[!htb]
\centering
\includegraphics[width=0.99\textwidth]{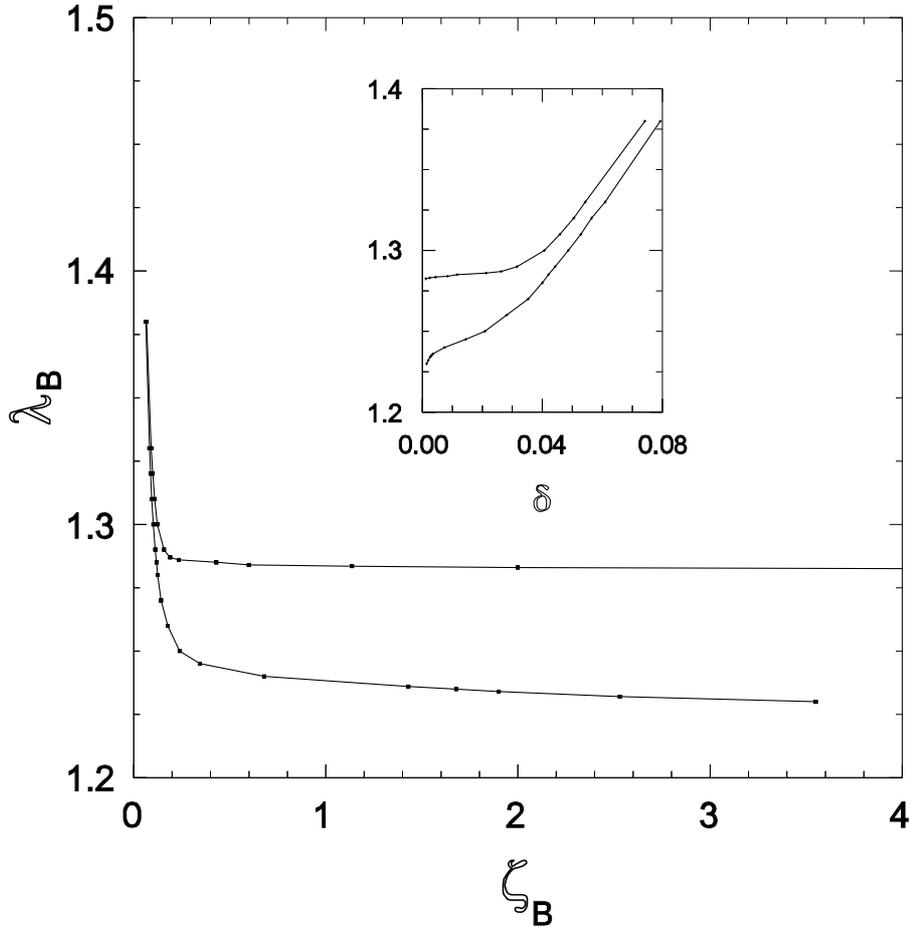}
\caption{Simulations in one dimension: SSS is observed in the region bounded by the two curves;
see text for parameters.  Inset: the same data plotted using $\delta = \zeta_A/\zeta_B$ instead of
$\zeta_B$ on the horizontal axis.}
\label{zmp200a}
\end{figure}

Figure \ref{zmp200a} shows the result of a systematic search in the $\zeta_B$-$\lambda_B$ plane on a ring of $L=200$ sites.  These data represent averages of
200 realizations, each lasting $10^5$ time units. SSS is observed in the region bounded by the two curves; the lower curve corresponds to $\tau_A = \tau_B$
while on the upper we have $\rho_A = \rho_B.$ There is an interval of $\lambda_B$ values, including $\lambda_c,$ on which SSS is observed for {\it
any} value of $\zeta_B$ greater than a certain minimum.  For larger values of $\lambda_B,$ SSS occurs only within a narrow range of $\zeta_B.$  In the latter
region we have $\alpha > \zeta_B > \zeta_A$, so that intraspecies competition is stronger than competition between species. Although we use a rather small
system to facilitate the search, SSS is not restricted to this system size.  For $L=800$ for example, the range of $\lambda_B$ values admitting SSS is
more restricted (from about 1.24 to 1.27) but at the same time the effect can be more dramatic. For $\lambda_B=1.27$ and $\zeta_B=0.2,$ for example, we find
$\rho_A=0.74,$ $\rho_B=0.49,$ and $\tau_B \approx 100 \tau_A.$ Studies using $\lambda_A = 1.35$ and $\zeta_A = 0.01$ yield a qualitatively similar diagram,
leading us to conjecture that the form of the region exhibiting SSS in one dimension is generically that shown in Fig. \ref{zmp200a}. To summarize, we verify
SSS in a limited but significant region of parameter space.

In the inset of Fig. \ref{zmp200a} the data are plotted in the $\delta$-$\lambda_B$ plane, for comparison with the theoretical predictions shown in Fig.
\ref{Phase_Diagram}. (Recall that parameter $g$ corresponds to the ratio $\lambda_B/\lambda_A,$ which is smaller than unity here.  The parameter $\delta =
\zeta_A/\zeta_B,$ i.e., the ratio of the interspecies competition rates.) The region exhibiting SSS is rather different from the predictions of both MFT and
stochastic field theory. In particular the lower boundary in the $\delta$-$\lambda_B$ plane is not horizontal and it is unclear whether the SSS region extends
to $\delta = 1.$  (As $\lambda_B$ is increased, SSS is observed in an ever more limited range of $\zeta_B$ values, making numerical work quite time-consuming.)
Here it is important to recall that irrelevant terms (in the renormalization group sense) are discarded in the theoretical analysis, so that predictions for the
phase diagram are only qualitative.

\begin{figure}[!htb]
\includegraphics[trim=1cm 1cm 6cm 15cm, clip=true]{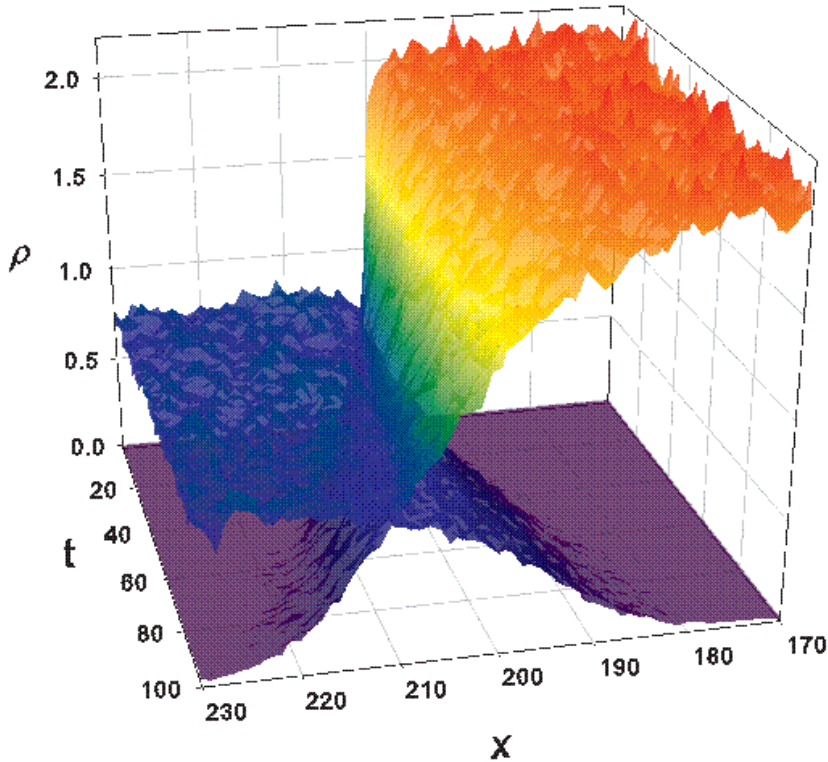}
\caption{Mean density profiles $\rho_A(x,t)$ (maximum at right) and $\rho_B(x,t)$ (maximum at left) starting
from neighboring single-species domains.  Parameters: $\lambda_A=1.6$, $\lambda_B=1.29$, $\zeta_A=0.005$,
$\zeta_B=0.16$.}
\label{cvspb}
\end{figure}

\begin{figure}[!htb]
\centering
\includegraphics[width=0.70\textwidth]{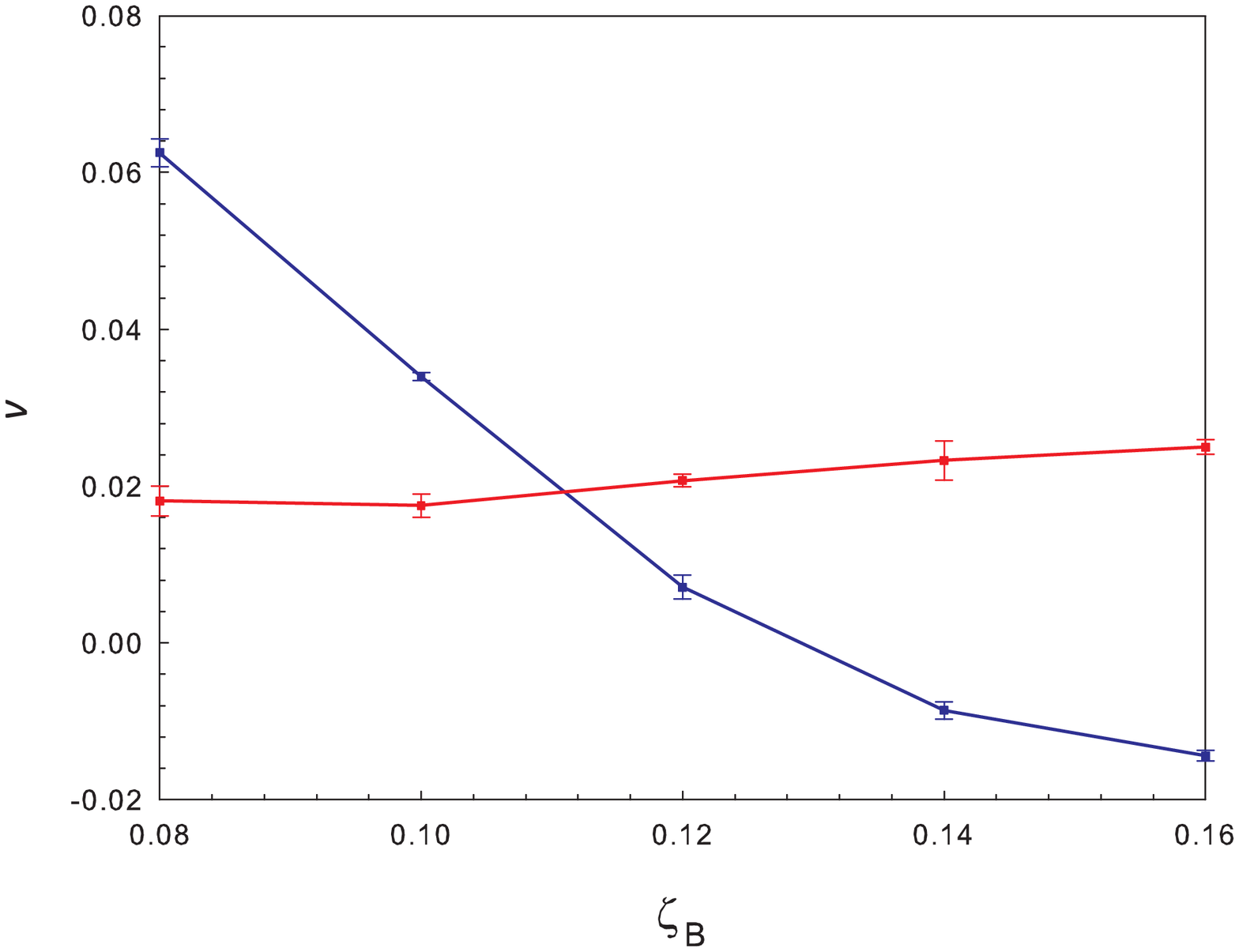}
\caption{Steady-state interface velocities for species A (blue) and B (red) versus $\zeta_B$; other
parameters as in Fig. \ref{cvspb}.}
\label{vfrpr}
\end{figure}

The evolution shown in Fig. \ref{ss5} suggests that survival of a given species depends on the speed of invasion into territory occupied by the competing
species; we study this issue via the following procedure.  We first use simulations to prepare collections of configurations drawn from the QS distribution for
each species in isolation. (We saved 1000 such configurations, on a ring of $L=200$ sites, for each $\lambda$ value of interest.)  To determine the speed of
invasion we prepare initial configurations for the two-species system by placing a randomly chosen, saved configuration with species A only at sites 1,...,L,
beside a similar configuration, with species B only, occupying sites $L+1,...,2L.$  In the subsequent dynamics, the two species interact at the
boundary, leading to invasion by one species or the other in different realizations. Fig. \ref{cvspb} shows the evolution of the A and B density profiles,
averaged over many realizations (3000 or more).  Given the average density profiles $\rho_A(x,t)$ and $\rho_B(x,t)$ we identify the interface position $X_i(t)$
of species $i$ via the condition $\rho_i[X_i(t),t] = \overline{\rho}_i/2,$ with $\overline{\rho}_i$ the QS density of species $i$ in isolation. Plotting the
interface positions versus time, we find that they attain steady velocities after a certain initial transient; the velocities are plotted versus $\zeta_B$ in
Fig. \ref{vfrpr}. (For the system size used here, steady velocities are attained after about 500 time units.) The interface velocity $v_B$ of species B
increases slowly with $\zeta_B$, but that of species A ($v_A$) falls rapidly, becoming negative for $\zeta_B$ greater than about 0.13.  For the parameters
considered here, the mean lifetime of the scarcer species (B) is longer for $\zeta_B \geq 0.113;$ very near this value, $v_A$ becomes smaller than $v_B.$  This
suggests that, as one would expect, the preferential extinction of the more populous species is associated with a smaller rate of spreading, so that A domains
eventually die out due to invasion by B.

\begin{figure}[!htb]
\centering
\includegraphics[trim=1cm 1cm 6cm 20cm, clip=true]{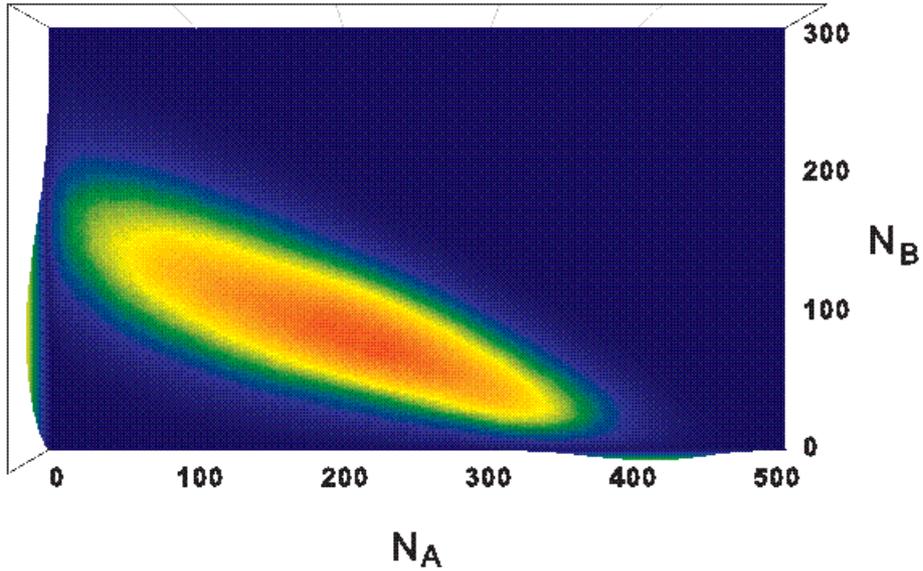}
\caption{Quasi-stationary probability distribution for parameters $\lambda_A=1.6$, $\lambda_B=1.26$, $\zeta_A=0.005$,
and $\zeta_B=0.3$, system size $L=200$.  Red corresponds to maximum and violet to zero probability.  The
maxima along the $x$ and $y$ axes correspond to the single-species QS distributions.}
\label{pqs}
\end{figure}

To close this discussion of simulation results, we present in Fig. \ref{pqs} a portrait of the QS probability distribution in the SSS region. The basic features
identified in the spatially uniform case (i.e., a well defined maximum corresponding to coexistence with $N_A > N_B,$ as well as single-species QS
distributions), persist in the presence of spatial structure.  We may speculate that the higher extinction probability for species A is associated with the more
diffuse barrier in the small-$N_A$ region (near the $y$-axis) as compared with the corresponding barrier near the $x$ axis.
More detailed simulations, including other parameter sets, larger systems, and simulations on the square lattice, are planned for future work.


\section{Conclusion}
\label{conclusion}

We study a spatial stochastic model of two-species competition, inspired by the spatially uniform system analyzed in Ref. \cite{survival_weaker}.  Based on
Janssen's results of for multicolored directed percolation \cite{janssen2001directed}, we find that the flow of parameters under the renormalization group is
such that a small competitive advantage may be amplified to the point of excluding the more numerous, but less competitive species, unless the latter reproduces
rapidly. Thus the less competitive population can go extinct, regardless of its size as given by mean-field theory. These effects tend to be stronger, the
smaller the dimensionality $d$ of the system. Our result shows that the survival of the scarcer phenomenon observed in \cite{survival_weaker} persists, in some
cases in a stronger form, in the presence of spatial structure. These differences are evident when comparing Figs. \ref{redner_fig} and \ref{Phase_Diagram}.

Monte Carlo simulations of a one-dimensional lattice model verify the existence of SSS when the more competitive species has a reproduction rate near the
critical value, while that of the less competitive one is above criticality. The latter species, although more numerous in the quasi-stationary regime, loses
out due to invasion by the more more competitive species.

Our results suggest that, in an ecological setting, a species with smaller population density can in fact out-compete one with a higher density,
even when the populations are not well mixed, so that the spatiotemporal pattern is one of shifting domains.  Under conditions that favor survival of the
scarcer, when both species are present in the same niche, the more competitive species (B) tends to survive longer, even if its density is smaller than the that
of the less competitive species (A). In isolation however, species A is {\it less} susceptible to extinction than is species B, again considered in isolation.
This suggests that, as long as both species are present globally, a cyclical dynamics of the form A $\to$ B $\to$ 0 $\to$ A,..., may be observed in localized
regions.  That is, a region occupied by species A is susceptible to invasion by B, which in turn is susceptible to extinction, permitting return of species A,
and so on.  Sequences of this kind are indeed seen in Fig. \ref{ss5}. Given the simplicity of the model and the limitations of the present theoretical and
numerical analysis, the above conclusions should be seen as preliminary at best.  They may nevertheless suggest directions for further research in ecological
modelling.

\section*{Acknowledgements}

This work is supported by CNPq, Brazil. \\

\bibliographystyle{nar.bst}
\bibliography{bibliografia.bib}
\end{document}